\documentclass[acmsmall]{acmart}

\usepackage[most]{tcolorbox}
\usepackage{xspace}
\usepackage{enumitem}
\usepackage{pifont}
\usepackage{multirow,makecell}
\usepackage{color}
\usepackage[table, dvipsnames]{xcolor}
\usepackage{listings,amsfonts}
\usepackage{caption}
\usepackage{subcaption}
\usepackage{threeparttable}
\usepackage{bbding}
\usepackage{graphicx}
\usepackage{booktabs} 
\usepackage{longtable}
\usepackage{flushend}
\usepackage{tabularx}
\usepackage{siunitx} 

\definecolor{best}{HTML}{FFEBEE} 
\definecolor{second}{HTML}{E3F2FD} 

\AtEndPreamble{
	\usepackage{hyperref}

}

\newcommand{\toolname}{\textsc{MalTotal}\xspace}

\newtcolorbox{snapshotbox}[1]{
    colback=black!4!white,
    colframe=black!12!white,
    boxrule=0.4pt,
    arc=2mm,
    boxsep=0pt,
    left=6pt, 
    right=6pt, 
    top=4pt, 
    bottom=5pt,
    fonttitle=\bfseries\sffamily,
    coltitle=black,
    title=#1,
    attach title to upper,
    after title={\par\vspace{1mm}}
}

\begin{document}
\title[\toolname: Cost-Effective and Language-Agnostic Malicious Code Poisoning Detection...]{\toolname: Cost-Effective and Language-Agnostic Malicious Code Poisoning Detection for Millions of Repositories}

\author[J. Zhao]{Jian Zhao}
\orcid{0009-0003-5716-1462}
\email{jian\_zhao\_@hust.edu.cn}
\authornotemark[1]
\affiliation{%
  \department{Hubei Key Laboratory of Distributed System Security}
  \department{Hubei Engineering Research Center on Big Data Security}
  \department{School of Cyber Science and Engineering}
  \institution{Huazhong University of Science and Technology}
  \city{Wuhan}           
  \country{China}
}

\author[S. Wang]{Shenao Wang}
\orcid{0000-0003-3818-3343}
\email{shenaowang@hust.edu.cn}
\authornote{Jian Zhao and Shenao Wang are the co-first authors.}
\affiliation{%
  \department{Hubei Key Laboratory of Distributed System Security}
  \department{Hubei Engineering Research Center on Big Data Security}
  \department{School of Cyber Science and Engineering}
  \institution{Huazhong University of Science and Technology}
  \city{Wuhan}           
  \country{China}
}

\author[Q. Wu]{Qingyang Wu}
\orcid{0009-0002-8279-3990}
\email{wuqingyang040802@gmail.com}
\affiliation{%
  \department{Hubei Key Laboratory of Distributed System Security}
  \department{Hubei Engineering Research Center on Big Data Security}
  \department{School of Cyber Science and Engineering}
  \institution{Huazhong University of Science and Technology}
  \city{Wuhan}           
  \country{China}
}

\author[Y. Zhao]{Yanjie Zhao}
\orcid{0000-0001-8793-5367}
\email{Yanjie\_Zhao@hust.edu.cn}
\authornote{Yanjie Zhao (yanjie\_zhao@hust.edu.cn) is the corresponding author.}
\affiliation{%
  \department{Hubei Key Laboratory of Distributed System Security}
  \department{Hubei Engineering Research Center on Big Data Security}
  \department{School of Cyber Science and Engineering}
  \institution{Huazhong University of Science and Technology}
  \city{Wuhan}           
  \country{China}
}

\author[X. Cheng]{Xiao Cheng}
\orcid{0000-0001-5456-3827}
\email{xiao.cheng@mq.edu.au}
\affiliation{%
  \institution{Macquarie University}
  \city{Sydney}
  \country{Australia}
}

\author[H. Wang]{Haoyu Wang}
\orcid{0000-0003-1100-8633}
\email{haoyuwang@hust.edu.cn}
\affiliation{%
  \department{Hubei Key Laboratory of Distributed System Security}
  \department{Hubei Engineering Research Center on Big Data Security}
  \department{School of Cyber Science and Engineering}
  \institution{Huazhong University of Science and Technology}
  \city{Wuhan}           
  \country{China}
}

\begin{abstract}
    The widespread adoption of open source software (OSS) has introduced significant security risks, with malicious code poisoning attacks increasingly targeting public package registries and open-source platforms. Existing detection approaches, including heuristic-, learning-, and LLM-based methods, suffer from language-specific designs, limited generalization, and high analysis costs, making them unsuitable for large-scale multi-language analysis. To address these challenges, we propose \textsc{MalTotal}, a scalable and cost-effective framework for language-agnostic malicious code detection. \textsc{MalTotal} leverages LLM-assisted semantic reasoning to identify sensitive APIs, perform hybrid semantic slicing, and reconstruct malicious behavior contexts while reducing analysis overhead. Our evaluations show that \textsc{MalTotal} outperforms 8 state-of-the-art baselines, achieving an average F1-score of 93.1\% across 5 mainstream languages. Its hybrid slicing reduces LLM token consumption by 94.0\%, lowering the analysis cost from \$86.25 to \$5.19 on 2,168 repositories. In a large-scale study of 120K GitHub repositories containing over 7.3 million files, \textsc{MalTotal} discovered 564 previously unknown malicious repositories across multiple languages at a total cost of \$338. These results demonstrate the effectiveness, scalability, and cost-efficiency of \textsc{MalTotal} in mitigating large-scale code poisoning attacks.
\end{abstract}



\maketitle

\section{Introduction}

The software supply chain has undergone a profound transformation over the past decade, with open source software (OSS) becoming a cornerstone of modern application development. OSS components now constitute up to 90\% of typical software applications~\cite{sonatype2024supplychain}, and ecosystems such as Node Package Manager (NPM)~\cite{npm} and Python Package Index (PyPI)~\cite{pypi} have experienced exponential growth. However, this widespread adoption has also introduced significant security risks~\cite{markus2019small,wermke2023contribute}. In particular, malicious code poisoning attacks~\cite{ohm2020knife,duan2021maloss} targeting public package registries have become a critical threat, leveraging techniques such as typosquatting~\cite{duan2021maloss,ladisa2023sokattack}, dependency confusion~\cite{birsan2021dependency,fossa2022dependency}, and vulnerability exploitation~\cite{zahan2022weaklinks,ladisa2023sokattack,gu2023package} to compromise downstream users.  Recent reports~\cite{sonatype2025malwareindex} identified 17,954 new open source malware packages in Q1 2025 alone, with more than 828,925 malicious packages discovered since 2019. These threats have further expanded beyond package managers to a broader range of OSS components, including VSCode extensions~\cite{vscodesecret,malvscode}, pre-trained models~\cite{zhao2024malmodel,zhu2025tensorflow}, MCP servers~\cite{poisonmcp,hou2025mcp}, and agent skills~\cite{poisonskill,david2025skill,maliciousskill}. Among them, GitHub has become a major attack surface due to its massive scale and central role in modern software development. Attackers exploit GitHub repositories through stealthy techniques such as repo confusion and fork bombs~\cite{reversinglabs2025banana,apiiro2025githubconfusion}, while real-world incidents such as the XZ Utils backdoor~\cite{akamai_xz_utils_backdoor,wikipedia_xz_utils_backdoor} demonstrate the severe impact of repository-level supply chain attacks. Recent studies further reveal the prevalence of this issue, with over 100,000 GitHub repositories affected in a large-scale repo confusion campaign~\cite{apiiro2025githubconfusion} and 9,294 out of 35.2K educational repositories containing malicious content~\cite{md2024educational}. Given that GitHub hosts over 518 million repositories~\cite{octoverse2024github}, detecting malicious code poisoning in large-scale, heterogeneous, and multi-language codebases remains a critical challenge.

\textbf{Research Gaps.}
To address these security concerns, researchers have proposed various detection techniques~\cite{ohm2023sokdec}, which can be broadly classified into heuristic-based~\cite{duan2021maloss,li2023malwukong,zheng2024oscar,cheng2024donapi}, learning-based~\cite{ladisa2023feasibility,sejfia2022practical}, and large language model (LLM)-based approaches~\cite{zahan2025socketai,di2024posterllm,wang2025malpacdetector}.
Heuristic-based methods rely on predefined rules or patterns, such as metadata~\cite{li2023malwukong,duan2021maloss} or malicious behavior patterns~\cite{gobbi2023codeql,cheng2024donapi}, and often use static~\cite{gobbi2023codeql,li2023malwukong,duan2021maloss} or dynamic analysis~\cite{zheng2024oscar,tanzir2025dysec} to detect threats~\cite{duan2021maloss,gobbi2023codeql}.
Learning-based approaches improve detection by learning malicious behavior patterns from labeled datasets, using features extracted from code~\cite{ladisa2023feasibility,simone2022feasibility,sejfia2022practical}, metadata~\cite{sajal2024metadata,haya2025mlpypi}, or behavior sequences~\cite{huang2024spiderscan,zhang2025killing,sun2024em4py}.
More recently, LLM-based methods have emerged as a promising direction, leveraging the reasoning capabilities of LLMs to analyze entire code files~\cite{zahan2025socketai,di2024posterllm} or package behaviors~\cite{wang2025malpacdetector,huang2024spiderscan}.
Despite these advancements, existing methods face significant challenges when applied to malicious code detection across millions of language-heterogeneous repositories:

\textit{L1: Limited Scope and Language-Specific Design.} Most methods are designed to work exclusively with specific programming languages (e.g., Python in PyPI~\cite{haya2025mlpypi,sun2024em4py,gao2024malguard} or JavaScript in NPM~\cite{huang2024spiderscan,wang2025malpacdetector,yu2024maltracker}), relying on ecosystem-specific behavior patterns~\cite{li2023malwukong,liang2023needle,cheng2024donapi} and manually curated sensitive API specifications~\cite{huang2024spiderscan}. Extending these tools to support additional languages often requires substantial re-engineering~\cite{li2023malwukong,zhang2025killing,ladisa2023feasibility,duan2021maloss}, which makes large-scale, multi-language detection across millions of repositories infeasible. Furthermore, attackers can bypass detection by leveraging sensitive APIs in third-party libraries~\cite{guo2023empirical,huang2024spiderscan}, resulting in inadequate coverage of malicious behaviors.

\textit{L2: Limited Generalization to Malicious Semantics.} Existing tools often abstract the rich and complex semantics of code into predefined behavioral patterns~\cite{duan2021maloss,li2023malwukong} or statistical features~\cite{ladisa2023feasibility,sajal2024metadata}, relying on pattern-matching against known malicious behaviors~\cite{cheng2024donapi,li2023malwukong} or classification through ML/DL models~\cite{ladisa2023feasibility,sejfia2022practical}. While these methods perform well for detecting known threats, they struggle with 0-day attack patterns or variations in malicious logic~\cite{vu2023badsnakes}. Additionally, learning-based methods typically yield binary classification results but offer little interpretable information to assist manual review~\cite{vu2023badsnakes,ladisa2023feasibility,sejfia2022practical}, thus burdening security teams.

\textit{L3: High Costs and Limited Scalability.} Emerging LLM-based approaches have yet to fully exploit LLMs' capabilities in analyzing malicious code semantics. Most methods either use LLMs for feature extraction~\cite{wang2025malpacdetector}, sensitive API analysis~\cite{huang2024spiderscan,gao2024malguard}, or data augmentation~\cite{yu2024maltracker}. Some approaches attempt to use LLMs as direct judges of malicious code, but they suffer from severe hallucination issues~\cite{zeshan2023silverbullet,ibiyo2025llmrag}, context window limitations~\cite{di2024posterllm,zahan2025socketai}, and expensive token processing costs~\cite{zahan2025socketai,di2024posterllm}. These limitations make it impractical to scale such methods across millions of repositories.

\textbf{Motivation \& Insights.}
Motivated by the limitations outlined in \textit{L1} and \textit{L2}, we propose LLMs as a silver bullet for language-agnostic malicious code semantic analysis. Our key insight is that LLMs, pre-trained on diverse code, possess domain knowledge on malicious patterns, and their alignment training enables them to distinguish malicious patterns from benign ones, making them a free lunch for language-agnostic malicious code detection. To validate this assumption, we conducted a pilot study where we used a few-shot approach to provide LLMs with repository-level, file-level, and minimal contextual code slices for evaluation. Our findings reveal that, with sufficiently simplified code contexts, LLMs can accurately reason about malicious code semantics.
However, applying the LLM-as-a-Judge paradigm to analyze millions of repositories introduces significant challenges stemming from \textit{L3}, specifically the context window limitations of LLMs and the prohibitive token costs associated with large-scale analysis. To address these issues, our research aims to design a scalable and cost-effective malicious semantic context slicing strategy for LLM-assisted repository-scale analysis. Inspired by the divide-and-conquer paradigm, we decompose this objective into two tasks: (1) parsing and identifying sensitive APIs across multi-language and third-party library environments, and (2) performing code slicing from sensitive API call sites while preserving malicious semantic contexts as much as possible.

\textbf{Our Work.}
In this paper, we propose \toolname, which uses LLMs for language-agnostic and cost-effective malicious code poisoning detection across millions of repositories. Specifically, \toolname first identifies sensitive APIs across multi-language and third-party libraries through LLM-assisted semantic summarization. This enables the discovery of derived sensitive operations in third-party libraries, with the summarization cached for efficient reuse. Building on this, \toolname performs backward code slicing for each identified sensitive operation, combining control dependency, data dependency, and call chain to construct a logically complete malicious behavior context. Finally, the sliced contexts are submitted to the LLM-as-a-Judge paradigm, which leverages its semantic reasoning capabilities to assess the intent of the code, classify potentially malicious behaviors, and provide interpretable reasons. By integrating these techniques, \toolname systematically addresses the scalability and token cost challenges while ensuring precise, language-agnostic malicious code detection across large-scale codebases.


\textbf{Contributions.}
To summarize, this paper makes the following contributions:
\begin{itemize}[leftmargin=15pt]
    \item \textbf{Language-Agnostic Analysis.} We propose \textsc{MalTotal}, a novel framework that leverages LLMs to enable language-agnostic malicious code detection. By identifying sensitive APIs across multi-language and third-party libraries, \textsc{MalTotal} generalizes beyond language-specific patterns and ecosystem-specific designs.
    
    \item \textbf{Cost-Effective Slicing.}  
    To address the token costs and scalability challenges of LLMs, \textsc{MalTotal} introduces a hybrid semantic context slicing strategy, reducing token consumption by 94.0\% while maintaining detection performance. This enables practical, cost-efficient analysis of large-scale repositories.
    
    \item \textbf{Scalability to Millions of Repositories.} 
    We demonstrate the scalability and practical impact of \textsc{MalTotal} by analyzing 120K GitHub repositories spanning over 7.3 million files. With a total cost of just \$338, \textsc{MalTotal} identified 564 previously unknown malicious repositories, showcasing its ability to handle real-world, large-scale codebases.
\end{itemize}
\section{A Pilot Study}
\label{sec:pilot_study}
To validate the feasibility of leveraging LLMs for malicious code detection, we conducted a pilot study to explore the capabilities and limitations of LLMs in analyzing malicious code semantics. This study aimed to explore two key questions: (1) Can LLMs accurately reason about malicious intent in code across diverse programming languages? (2) What are the practical challenges, such as context window limitations and token consumption, when scaling to millions of repositories?

\subsection{Study Setup}

\noindent \textbf{Dataset.}
To evaluate the feasibility of LLM-assisted malicious code detection, we curated malicious code samples from MalwareBench~\cite{li2025malwarebench} (Python and JavaScript) and SourceFinder~\cite{md2020sourcefinder} (Java, Go, and PHP). To ensure diversity, we stratified projects by size (LoC and file counts) into five tiers and randomly sampled 10 valid projects from each tier for each language, resulting in 50 samples per language. Each sample was manually reviewed and annotated by three independent reviewers to verify its malicious behavior, with invalid or duplicate samples replaced by new samples from the same tier. The resulting dataset covers diverse malicious behaviors, including backdoors, spyware, sniffers, and cryptomining. We focus only on malicious samples in this pilot study, as our goal is to evaluate whether LLMs can effectively understand malicious code semantics rather than distinguish malicious code from benign code.

\noindent\textbf{Code Context Configurations.}
To investigate the influence of different levels of contextual information on LLM-assisted malicious code detection, we prepared three granular levels of input contexts for each malicious sample.
\begin{itemize}[leftmargin=15pt]
    \item \textbf{Repository-Level (Repo):} The full source code of the repository, concatenated into a single input context. This strategy is similar to SocketAI~\cite{zahan2025socketai}, which provides the most complete context but often exceeds the model's context window and incurs high token costs. 
    \item \textbf{File-Level (File):} All files within the repository that contain sensitive API calls, concatenated into a single input context. This configuration balances context richness and token cost.
    \item \textbf{Slice-Level (Slice):} A manually curated, minimal code snippet that represents the complete malicious logic extracted from the repository. By eliminating irrelevant or unrelated code, this configuration represents the idealized minimum context required for analyzing malicious logic.
\end{itemize}

\noindent\textbf{LLM and Prompting Strategy.}
We used DeepSeek-V3~\cite{deepseekai2024deepseekv3technicalreport} as the foundational LLM in our pilot study. DeepSeek-V3 was selected due to its reasoning capabilities and relatively lower computational cost, making it a practical choice for this study. While it is not the current SOTA model for code reasoning, its performance is representative of trends expected from other advanced LLMs under similar conditions. The detailed comparison of different LLMs is provided in \autoref{sec:rq3}.
For the prompting strategy, we employed commonly used techniques such as role assignment, Chain-of-Thought~(CoT) guidance, few-shot learning, and structured output to guide the model's reasoning process. The specific design of these prompts is described in \autoref{sec:judge}.

\begin{table}[t]
\centering
\fontsize{8}{11}\selectfont
\caption{Performance comparison of different context levels across languages.}
\label{tab:reliability_cost_comparison}
\begin{tabular}{cccccrrrr}
\hline
\textbf{Language} & \textbf{Context} & \textbf{\#All} & \textbf{Avg. LoC} & \textbf{Avg. \#Files} & {\textbf{OOC$^\ast$ (\%)}} & {\textbf{FNR (\%)}} & \textbf{Tokens} & \textbf{Cost$^\dag$} \\
\hline
\multirow{3}{*}{\textbf{JavaScript}} & Repo & \multirow{3}{*}{50} & \multirow{3}{*}{1085} & \multirow{3}{*}{9.5} & 14.0 & 16.3 & 1,502,941 & \$0.41 \\
& File & & & & 4.0 & 14.6 & 900,943 & \$0.24 \\
& \textbf{Slice} & & & & \textbf{0.0} & \textbf{2.0} & \textbf{86,432} & \textbf{\$0.02} \\
\hline
\multirow{3}{*}{\textbf{Python}} & Repo & \multirow{3}{*}{50} & \multirow{3}{*}{1772} & \multirow{3}{*}{6.3} & 40.0 & 0.0 & 2,141,938 & \$0.58 \\
& File & & & & 32.0 & 0.0 & 865,686 & \$0.23 \\
& \textbf{Slice} & & & & \textbf{0.0} & \textbf{0.0} & \textbf{133,212} & \textbf{\$0.04} \\
\hline
\multirow{3}{*}{\textbf{Go}} & Repo & \multirow{3}{*}{50} & \multirow{3}{*}{6428} & \multirow{3}{*}{28.3} & 14.0 & 0.0 & 5,353,978 & \$1.45 \\
& File & & & & 0.0 & 0.0 & 338,074 & \$0.09 \\
& \textbf{Slice} & & & & \textbf{0.0} & \textbf{0.0} & \textbf{128,943} & \textbf{\$0.03} \\
\hline
\multirow{3}{*}{\textbf{Java}} & Repo & \multirow{3}{*}{50} & \multirow{3}{*}{3888} & \multirow{3}{*}{23.2} & 12.0 & 2.3 & 2,515,472 & \$0.68 \\
& File & & & & 0.0 & 2.0 & 309,808 & \$0.08 \\
& \textbf{Slice} & & & & \textbf{0.0} & \textbf{0.0} & \textbf{69,991} & \textbf{\$0.02} \\
\hline
\multirow{3}{*}{\textbf{PHP}} & Repo & \multirow{3}{*}{50} & \multirow{3}{*}{7704} & \multirow{3}{*}{12.1} & 46.0 & 3.7 & 15,633,522 & \$4.22 \\
& File & & & & 12.0 & 2.3 & 2,855,182 & \$0.77 \\
& \textbf{Slice} & & & & \textbf{0.0} & \textbf{0.0} & \textbf{107,396} & \textbf{\$0.03} \\
\hline
\multirow{3}{*}{\textbf{Total}} & \textbf{Repo} & \multirow{3}{*}{250} & \multirow{3}{*}{4176} & \multirow{3}{*}{15.9} & \textbf{25.2} & \textbf{4.1} & \textbf{39,703,851} & \textbf{\$10.72} \\
& \textbf{File} & & & & \textbf{9.6} & \textbf{3.4} & \textbf{8,182,693} & \textbf{\$2.21} \\
& \textbf{Slice} & & & & \textbf{0.0} & \textbf{0.5} & \textbf{939,916} & \textbf{\$0.25} \\
\hline
\end{tabular}
\begin{tablenotes}
\small
\item \textbf{$\ast$ OOC~(Out-of-Context)(\%):} The percentage of samples where the input exceeded the maximum context, causing analysis failures.
\item \textbf{$\dag$ Cost:} The cost is calculated based on the official DeepSeek pricing of \$0.27 per million tokens.
\end{tablenotes}
\end{table}

\subsection{General Findings}
Our experimental results, as summarized in ~\autoref{tab:reliability_cost_comparison}, provide a comprehensive evaluation of the performance, cost, and scalability of LLMs for malicious code detection. To address the two key questions outlined in our study, we summarize the findings as follows:

\noindent\textbf{Accuracy Across Context Levels.}  
Our results demonstrate that LLMs can effectively identify malicious intent across programming languages when provided with optimized context granularity. The \textit{Slice Context}, which focuses on minimal, directly relevant code snippets, achieved the lowest false negative rate (FNR) across all languages (e.g., 2.0\% for JavaScript and 0.0\% for Python), indicating that LLMs can accurately reason about malicious behaviors when unnecessary noise is removed. However, the \textit{Repository Context}, while comprehensive, suffers from higher FNRs in some cases (e.g., 16.3\% in JavaScript) due to irrelevant or excessive information. The \textit{File Context}, serving as a middle ground, provides reasonable accuracy while reducing token consumption. These findings confirm that LLMs are promising for malicious code detection across diverse languages, especially when the input context is carefully curated.

\noindent\textbf{Scalability and Cost Efficiency.}  
The \textit{Repository} and \textit{File Contexts} face significant challenges due to context window limitations, which restrict the ability of LLMs to process large codebases effectively. As shown in ~\autoref{tab:reliability_cost_comparison}, 46.0\% of PHP samples and 40.0\% of Python samples could not be analyzed at the \textit{Repository Context}, resulting in a high percentage of analysis failures. Although the \textit{File Context} mitigates this issue to some extent, it still struggles with larger files. In contrast, the \textit{Slice Context} completely eliminates token window constraints, achieving 0.0\% N.A. samples across all languages by focusing on minimal, relevant code snippets.
In addition to these technical constraints, the choice of context granularity has a profound impact on analysis cost. Analyzing 409 samples at the \textit{Repository Context} consumed approximately 39.7 million tokens, costing \$10.72, while the \textit{File Context} required 8.2 million tokens (\$2.21). By comparison, the \textit{Slice Context} consumed only 0.94 million tokens, costing just \$0.25, which reaches a 97.6\% reduction in both token consumption and cost relative to the \textit{Repository Context}. When scaled to millions of repositories, these differences become even more pronounced: repository-level analysis is estimated to cost over \$26K for one million repositories, compared to just \$625 for the \textit{Slice Context}. While this estimation is approximate, it provides a general indication of the significant computational and cost burden for large-scale deployment. These results underscore the critical importance of optimizing context granularity to ensure cost-effective and scalable large-scale deployment.

\noindent \textbf{Motivation and Insights.}
The results of this pilot study provide valuable insights that motivate the design of \toolname. While LLMs demonstrate great promise as language-agnostic detectors for identifying malicious code, directly applying them to vast raw codebases presents significant scalability and cost challenges. Our findings highlight that the core issue lies in providing LLMs with minimal yet precise malicious code contexts, as excessive or irrelevant information can lead to inefficiencies and reduced accuracy. These insights underscore the need for an automated framework that can extract precise malicious code contexts efficiently and at scale.
\begin{figure}[t]
    \centering
    \includegraphics[width=0.85\linewidth]{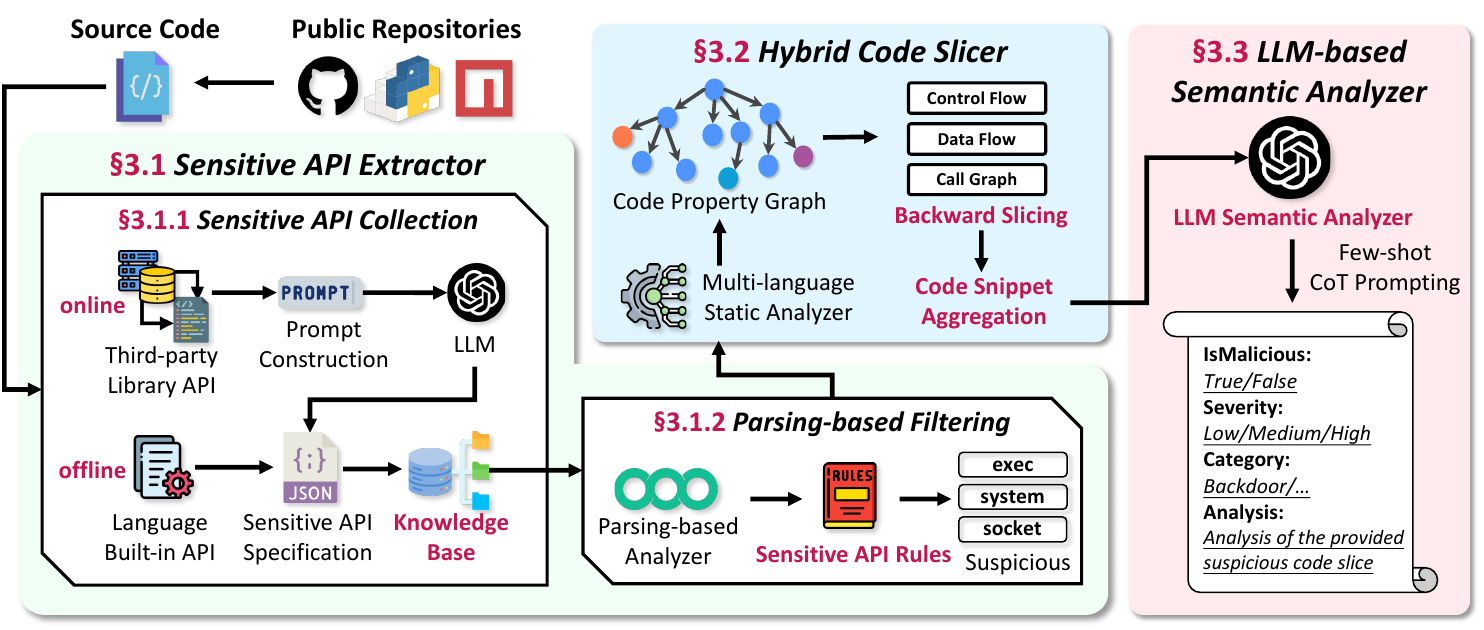}
    \caption{The workflow of \toolname}
    \label{fig:architecture}
\end{figure}

\section{Design of \toolname}
\label{sec:methodology}

To address the challenges of language-agnostic, scalable, and cost-effective malicious code detection, we present \toolname, as shown in ~\autoref{fig:architecture}. The methodology consists of three key components: \textit{Sensitive API Extractor}, \textit{Hybrid Code Slicer}, and \textit{LLM-based Semantic Analyzer}. 


\subsection{Sensitive API Extractor}
\label{sec:identifier}

Drawn from existing studies~\cite{li2023malwukong,cheng2024donapi,huang2024spiderscan}, almost all malicious behaviors across programming languages are typically implemented through a series of sensitive APIs and their combined sequences of operations, such as those enabling code execution, cryptographic operations, and file operations. Rule-based methods and learning-based methods have widely utilized these sensitive API sequences to model malicious behaviors~\cite{li2023malwukong,cheng2024donapi,huang2024spiderscan}.
Following this insight, we also take the identification of sensitive APIs as the first step in pinpointing potential malicious behaviors within large codebases.
However, prior works have primarily defined sensitive APIs based on heuristic knowledge, and they are often limited to built-in functions in Python and JavaScript~\cite{li2023malwukong,zheng2024oscar,duan2021maloss,gao2024malguard}. The most relevant work, SpiderScan~\cite{huang2024spiderscan}, extends the sensitive API specification by considering TPLs within the NPM ecosystem. However, it is constrained to JavaScript and only reports the number of sensitive TPL APIs identified, without providing a usable or comprehensive list. 
To address these limitations, \toolname systematically identifies sensitive APIs through two steps. First, \toolname collects and organizes sensitive APIs, including both built-in functions and TPL functions, by leveraging LLM-assisted semantic analysis. Next, it employs parsing-based static analysis to preliminarily locate and identify sensitive operations within large codebases.

\subsubsection{Sensitive API Collection}
To comprehensively identify sensitive behaviors within codebases, \toolname systematically collects sensitive APIs by targeting two main sources: \textit{language-specific built-in APIs} and \textit{third-party APIs} dynamically identified from external libraries. The collection of sensitive APIs is divided into two distinct phases: \textit{language-specific built-in APIs} are collected through an \textit{offline} process, where a static knowledge base is constructed prior to the analysis phase; in contrast, \textit{third-party APIs} are identified and analyzed in an \textit{online} manner, where LLMs are employed during runtime to analyze and summarize the semantics of third-party APIs.
As summarized in ~\autoref{tab:sensitive_behavior_taxonomy}, these APIs are categorized based on their potential risk and malicious usage patterns, using a heuristically constructed taxonomy derived from previous rule-based and learning-based methods~\cite{li2023malwukong,cheng2024donapi,huang2024spiderscan}. Specifically, the taxonomy consists of five major categories: payload execution, network access, file operations, sensitive data access, and cryptography.

\begin{table}[t]
\centering
\caption{Taxonomy of sensitive APIs and representative examples across languages.}
\label{tab:sensitive_behavior_taxonomy}
\resizebox{\linewidth}{!}{%
\begin{tabular}{lll}
\hline
\textbf{Category} & \textbf{Subcategory} & \textbf{Representative API Examples} \\
\hline
\multirow{3}{*}{\textbf{Payload Execution}} 
& System Command Execution & \texttt{subprocess.run} (L1), \texttt{exec} (L1, L5) \\
& Dynamic Code Execution & \texttt{eval}, \texttt{exec} (L1, L5), \texttt{javax.script.ScriptEngineManager} (L4) \\
& External File Execution & \texttt{subprocess.run} (L1), \texttt{proc\_open} (L5) \\
\hline
\multirow{4}{*}{\textbf{Network Access}} 
& Create Connections/Servers & \texttt{socket.bind/listen} (L1, L5), \texttt{net.Listen} (L3) \\
& Resolve DNS & \texttt{dns.lookup} (L2), \texttt{dns\_get\_record} (L5) \\
& Send Data & \texttt{socket.send} (L1, L5), \texttt{conn.Write} (L3) \\
& Receive Data & \texttt{socket.recv} (L1, L5), \texttt{conn.Read} (L3) \\
\hline
\multirow{6}{*}{\textbf{File Operation}} 
& Create or Delete Files & \texttt{os.mkdir} (L1), \texttt{mkdir} (L5), \texttt{os.Mkdir} (L3) \\
& Read Files & \texttt{open} (L1, L5), \texttt{file\_get\_contents} (L5), \texttt{os.Open} (L3) \\
& Write to Files & \texttt{os.write} (L1), \texttt{file\_put\_contents} (L5), \texttt{os.Create} (L3) \\
& Modify File Permissions & \texttt{os.chmod} (L1, L5), \texttt{os.Chmod} (L3) \\
& Link Operations & \texttt{os.symlink} (L1), \texttt{symlink} (L5), \texttt{os.Symlink} (L3) \\
& Search Files or Directories & \texttt{os.walk} (L1), \texttt{scandir} (L5), \texttt{filepath.Walk} (L3) \\
\hline
\multirow{3}{*}{\textbf{Sensitive Data Access}} 
& System Information & \texttt{os.uname} (L1), \texttt{php\_uname} (L5), \texttt{runtime.GOOS} (L3) \\
& Env/Process Information & \texttt{os.getenv} (L1, L5), \texttt{process.env} (L2), \texttt{os.Getenv} (L3) \\
& User Information & \texttt{getpass.getuser} (L1), \texttt{get\_current\_user} (L5) \\
\hline
\multirow{3}{*}{\textbf{Cryptography}} 
& Create Cipher Objects & \texttt{rsa.generate\_private\_key} (L1) \\
& Encryption/Decryption & \texttt{crypto.createCipheriv} (L4), \texttt{openssl\_encrypt} (L5) \\
& Encoding/Decoding & \texttt{base64.b64encode} (L1), \texttt{base64\_encode} (L5) \\
\hline
\end{tabular}%
}
\begin{tablenotes}
\fontsize{8}{10}\selectfont
\item $\ast$ \textbf{L1}: Python, \textbf{L2}: JavaScript, \textbf{L3}: Go, \textbf{L4}: Java, \textbf{L5}: PHP.
\end{tablenotes}
\end{table}

\noindent\textbf{Language-specific Built-in APIs.}  
To construct a comprehensive knowledge base of built-in sensitive APIs, we integrated three primary sources: MalOSS, MalTracker, and official language documentation. First, we consolidated sensitive API specifications from MalOSS~\cite{duan2021maloss}, which provides a detailed list of 1,257 sensitive APIs across Python, JavaScript, Java, and PHP. To further expand the knowledge base, we integrated 293 additional sensitive APIs for JavaScript extracted from MalTracker~\cite{yu2024maltracker}, ensuring no overlap with MalOSS. Next, we manually curated sensitive APIs from official language documentation~\cite{jsdocs,pythondocs,godocs,javadocs,phpdocs}, especially focusing on Go that no prior specifications were available. This process contributed 597 new sensitive APIs across all supported languages, including augmentations for Python, JavaScript, Go, Java, and PHP. As shown in \autoref{tab:api_statistics}, after combining all three sources and removing duplicates, we offline constructed a comprehensive sensitive API knowledge base comprising 2,147 built-in APIs\footnote{The complete list of these 2,147 sensitive built-in APIs is publicly available in our \href{https://figshare.com/s/43a79dbd719dbda407f4}{Anonymous Artifact}.}.

\begin{table}[t]
\centering
\caption{Statistics of collected built-in and TPL sensitive APIs across languages.}
\label{tab:api_statistics}
\fontsize{8}{10}\selectfont 
\setlength{\tabcolsep}{3pt} 
\begin{tabular}{|c|rr|rr|rr|rr|rr|rr|}
\hline
\multirow{2}{*}{\textbf{Language}} & \multicolumn{2}{c|}{\textbf{Execution}} & \multicolumn{2}{c|}{\textbf{Network}} & \multicolumn{2}{c|}{\textbf{File Operation}} & \multicolumn{2}{c|}{\textbf{Sensitive Data}} & \multicolumn{2}{c|}{\textbf{Cryptography}} & \multicolumn{2}{c|}{\textbf{Total}} \\ 
\cline{2-3} \cline{4-5} \cline{6-7} \cline{8-9} \cline{10-11} \cline{12-13}
 & \textbf{Built-in} & \textbf{TPL} & \textbf{Built-in} & \textbf{TPL} & \textbf{Built-in} & \textbf{TPL} & \textbf{Built-in} & \textbf{TPL} & \textbf{Built-in} & \textbf{TPL} & \textbf{Built-in} & \textbf{TPL} \\ 
\hline
Python      & 127 & 52 & 153 & 97 & 141 & 58 & 14 & 4  & 23 & 0 & 458 & 211 \\ 
JavaScript  & 79 & 31 & 149 & 41  & 194 & 59 & 20 & 14  & 82 & 2 & 524 & 147 \\ 
Go          & 22 & 17  & 45  & 63   & 58  & 30  & 34 & 42  & 55 & 5  & 214 & 157   \\ 
Java        & 80 & 1  & 225 & 26   & 172 & 7  & 19 & 1  & 175 & 0  & 671 & 35   \\ 
PHP         & 69 & 26  & 63  & 37   & 84  & 26  & 3  & 20  & 61 & 35  & 280 &  144  \\ 
\hline
\textbf{Total} & \textbf{377} & \textbf{101} & \textbf{635} & \textbf{227} & \textbf{649} & \textbf{154} & \textbf{90} & \textbf{61} & \textbf{396} & \textbf{7} & \textbf{2147} & \textbf{694} \\ 
\hline
\end{tabular}
\end{table}

\begin{figure}[t]
    \centering
    { 
    \sffamily\fontsize{7pt}{10.5pt}\selectfont

    \begin{snapshotbox}{System Prompt}
        You are a security expert. Your task is to analyze the provided \{language\} third-party package API for potential malicious usage.

        \#\#\# TASK:
        Analyze the API and determine whether it involves any of the following sensitive behaviors:
        1. Payload execution: ...
        2. Network access: ...
        3. File operations: ...
        4. Sensitive data access: ...
        5. Cryptography: ...

        \#\#\# INPUT FORMAT:
        I will provide the following information:
        1. Package name: \{package\_name\}
        2. Version: \{package\_version\}
        3. API path: \{api\_path\}
        4. Object type: \{object\_type\}
        5. Source code: \{source\_code\}

        \#\#\# OUTPUT FORMAT: 
        Return the analysis in JSON format with the following fields:
\begin{verbatim}
{
    "is_risky": <Boolean>, 
    "risk_types": [<String array of sensitive behaviors>], 
    "description": "<Detailed explanation>", ...
}
\end{verbatim}
        \#\#\# SPECIAL INSTRUCTIONS:
        1. Identify specific code snippets or function calls that exhibit sensitive behaviors.
        2. Use the five categories of sensitive behaviors to classify risks...

    \end{snapshotbox}
    
    } 
    \caption{The partial prompt template used for LLM-based third-party API analysis across languages.}
    \label{fig:prompt_template}
\end{figure}

\noindent\textbf{Third-party Library APIs.}  
The analysis of TPLs is conducted as an \textit{online process}, tailored to the dependencies of the target project. During analysis, \toolname first parses the dependency configuration files of the project (e.g., \texttt{pyproject.toml} for Python, \texttt{package.json} for JavaScript, etc.) to enumerate all declared TPLs and their corresponding versions. These libraries are then downloaded from their respective official repositories for analysis.  
To identify third-party APIs, \toolname statically analyzes the project’s source code to locate all imported functions and resolve their origins by inspecting their namespaces. If a function is determined to originate from a TPL, \toolname retrieves its implementation by resolving the library structure and locating the corresponding source files. To maintain cost-effectiveness and prevent token bloat, \toolname strictly restricts the extraction scope to the API's immediate function body, without recursively tracing its transitive callees. In practice, the function signature, its immediate implementation, and any available inline documentation provide sufficient semantic cues for intent inference, a design consistent with prior scalable analyses like SpiderScan\cite{huang2024spiderscan}. The extracted code for each function is then provided to an LLM, which performs semantic reasoning to analyze its behavior and determine whether it corresponds to a sensitive operation, such as payload execution, network access, file operation, sensitive data access, and cryptography. 
Specifically, the semantic reasoning process is guided by a prompt template, which instructs the LLM to generate structured JSON outputs with risk assessments and specific reasons. A partial version of this prompt template is shown in \autoref{fig:prompt_template}.
The structured semantic summary generated for each third-party library API, regardless of whether it is identified as sensitive, is cached along with the (library, version, function) tuple. By caching these summaries, \toolname avoids redundant reanalysis of the same APIs across different projects, improving efficiency and scalability.
As part of our evaluation on the \textit{Multi-Lang-Bench} (see \autoref{sec:setup}), \toolname has analyzed and cached a total of 27,746 third-party APIs, among which 694 are sensitive APIs\footnote{The complete list of these 694 sensitive third-party APIs is publicly available in our \href{https://figshare.com/s/43a79dbd719dbda407f4}{Anonymous Artifact}.}. These sensitive third-party APIs are categorized and summarized in \autoref{tab:api_statistics}, showcasing the diversity of sensitive operations introduced by TPLs across different ecosystems. 

\subsubsection{Parsing-based Filtering}  
To efficiently filter suspicious projects and avoid unnecessary heavy analysis, we employ a parsing-based multi-language analyzer, Semgrep~\cite{semgrep}, as a lightweight pre-processing step. This process aims to quickly determine whether a target project contains any sensitive API usages. If no sensitive APIs are identified, subsequent slicing and analysis are skipped, reducing the computational overhead while analyzing large-scale repositories.

\noindent\textbf{Direct Calls.}  
Direct calls refer to explicit invocations of sensitive APIs, such as \texttt{os.system} in Python or \texttt{exec} in JavaScript. For each sensitive API in the knowledge base, we compile specific rules to match these direct patterns, allowing us to efficiently identify sensitive API usage.

\noindent\textbf{Indirect Calls and Aliases.}  
Indirect calls occur when sensitive functions are invoked through intermediate references or aliases, such as \texttt{import os.system as executor} or \texttt{executor = os.system} in Python. For these cases, we intentionally avoid using alias analysis due to the high computational cost of tracking the point-to relationships at the pre-processing stage. Instead, we focus on identifying potential suspicious projects by matching sensitive library imports and alias assignment statements, which ensures that our pre-processing remains scalable and efficient.

By scanning the codebase, we identify all occurrences of sensitive APIs or libraries, either through direct calls or indirect references. Each matched instance is recorded with its corresponding location, API name, category, and source code for further processing. Projects without any matched sensitive APIs are filtered out, allowing us to focus subsequent slicing and deeper analysis on projects that are more likely to exhibit sensitive behaviors.

\subsection{Hybrid Code Slicing}
\label{sec:slicer}
The hybrid code slicer extends the sensitive APIs identified during the parsing stage into semantically rich and logically complete \textit{Suspicious Behavior Contexts} for further analysis. \toolname's slicer is implemented using \textit{Joern}~\cite{joern}, a fully open-source static analyzer that supports Code Property Graph~(CPG)~\cite{code_property_graph}. While tools such as CodeQL~\cite{codeql} are also capable of performing code slicing, we choose Joern for its fully open-source nature and ease of extension, which make it well-suited for our workflow. Nevertheless, \toolname is generalizable and can be extended to other multi-language static analyzers. 

\noindent \textbf{CPG Generation and Pre-processing.}  
The slicing process begins with the generation of a Code Property Graph~(CPG), a unified intermediate representation that combines the Abstract Syntax Tree~(AST), Control Flow Graph~(CFG), and Program Dependence Graph~(PDG) into a language-agnostic graph structure. Using Joern, we convert the source code of each target project into a corresponding CPG, which serves as the foundation for subsequent slicing operations. For sensitive APIs identified during the filtering stage, we distinguish between two scenarios: direct calls and alias-based references. For direct calls, we initiate a backward slicing from the CPG node of the API call to identify all relevant ancestor code that could influence its execution. For alias-based references, where sensitive APIs are accessed via intermediate aliases, we perform a forward data-flow analysis starting from the alias definition node, such as an alias assignment or import statement. This forward trace identifies the actual call sites where the alias invokes the sensitive API. Once these call sites are located, we apply backward slicing from these positions to extract their complete contexts of execution. 

\begin{figure}[t]
    \centering
    \includegraphics[width=\linewidth]{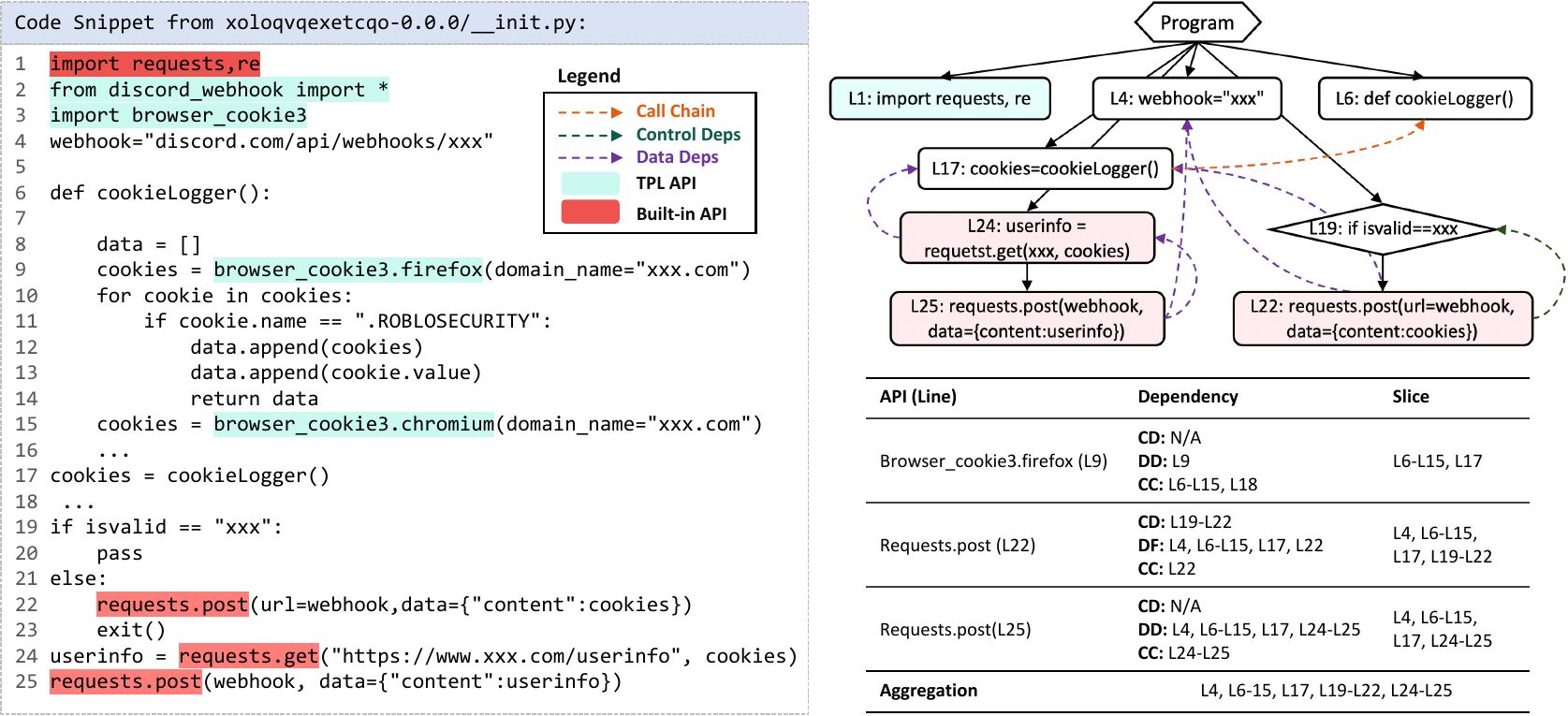} 
    \caption{An example of malicious behavior context extraction using hybrid slicing on the CPG. It includes (1) the input code snippet, (2) the CPG with call, control, and data dependencies, and (3) the aggregated slice that captures cookie collection and data exfiltration behavior.}
    \label{fig:example}
\end{figure}

\noindent\textbf{Backward Hybrid Slicing.}  
Following prior taint-based malicious package detection, we observe that malicious behaviors typically culminate in sensitive operations, i.e., security-relevant sinks such as data exfiltration or command execution. Based on this observation, \toolname adopts a sink-driven backward slicing approach. This design mitigates the path explosion often caused by forward slicing from data sources, because a source value may propagate through numerous aliases, branches, function calls, and intermediate variables, most of which do not eventually reach a security-relevant sink. By starting from sinks, \toolname focuses only on the control and data dependencies that directly affect sensitive operations, thereby reducing irrelevant paths and code contexts. Specifically, for each sensitive API call, whether identified directly or resolved through alias tracing, we recursively traverse the CPG to extract all ancestor code that could potentially influence its behavior. This slicing process combines three types of dependencies: call-chain, control dependency, and data dependency slicing, allowing us to capture a complete representation of the suspicious behavior context. To prevent context bloat during this call-chain expansion, \toolname enforces a maximum call depth constraint (default $k=3$) and includes the whole callee function body. While strict intra-procedural slicing extracts fewer lines, preserving the complete structural context of the callee is crucial for LLMs to accurately comprehend the semantics without losing critical contextual cues, as empirically validated in \autoref{sec:RQ2 Ablation Study}. For example, as shown in \autoref{fig:example}, the \texttt{cookieLogger} function (L6) collects browser cookies using the \texttt{browser\_cookie3} library (L9 and L15), and the collected cookies are subsequently exfiltrated to a hardcoded webhook URL via two separate \texttt{requests.post} calls (L22 and L25). 
When analyzing the API call \texttt{browser\_cookie3.firefox} within the \texttt{cookieLogger} function (L9), call-chain slicing is used to expand the context to include the entire \texttt{cookieLogger} function and its call site (L6-L15 and L17). For the \texttt{requests.post} call located in the \texttt{else} branch of the conditional block (L22), control dependency slicing captures the enclosing \texttt{if-else} block (L19-L22) that dictates its execution conditions. Finally, data dependency slicing on both \texttt{requests.post} calls identifies their shared data origins, such as the \texttt{webhook} variable (L4) and the \texttt{cookieLogger} function, providing a complete context for the suspicious behavior. For example, the second \texttt{requests.post} call (L25) depends on user information fetched via \texttt{requests.get} (L24) and shares the same webhook destination. The raw output of this stage includes slices for each sensitive API call, capturing the relevant lines of code and their dependencies.

\noindent\textbf{Context Refinement and Aggregation.}  
The raw slices generated often contain significant redundancies and overlaps, making them inefficient for direct LLM analysis. To address this, we perform a two-step refinement process. First, we apply maximal content filtering, where any slice that is a proper subset of another is discarded. For instance, in \autoref{fig:example}, the raw slice generated for \texttt{browser\_cookie3.firefox} is entirely contained within the larger slices generated for the \texttt{requests.post} calls and is therefore discarded. Next, we perform similarity-based iterative aggregation, where slices whose line-level Jaccard similarity (the ratio of intersecting lines to the union of lines) exceeds a predefined empirical threshold ($\tau = 0.3$ in our implementation) are merged into a single, more holistic context. Crucially, this empirical threshold prevents the aggregation from degenerating into a naive ``merge-by-line'' union. Semantically distinct slices that merely share trivial statements (e.g., common \texttt{import} declarations or global variables) will yield a low Jaccard similarity and remain separate. For example, the two slices for the \texttt{requests.post} calls, while not identical, share substantial code related to cookie theft and data exfiltration. These slices are aggregated into a unified context that encapsulates the entire data exfiltration routine (L4, L6-L15, L17, L19-L22, and L24-L25). This merging process continues iteratively until no more slices satisfy the similarity condition, resulting in a minimal set of highly condensed, logically distinct contexts. These refined slices are optimized for the final stage of semantic analysis, ensuring both efficiency and completeness in analyzing malicious behaviors.

\begin{figure}[t!]
    \centering
    { 
    \sffamily\fontsize{7pt}{10.5pt}\selectfont

    \begin{snapshotbox}{System Prompt}
        You are a security expert. I will provide a code slice and you must determine if it is malicious.
        
        \#\#\# TASK:
        Your goal is to analyze the provided code snippets and determine if they contain malicious behavior. To do this, follow these steps in your reasoning process:
        1. Identify sensitive operations: Pinpoint any code that performs actions like ...
        2. Trace data flow: Trace the origin and flow of data to these sensitive operations ...
        3. Assess intent: Based on the data flow and the context, determine...
        4. Synthesize Findings: Consolidate your findings into the final JSON output.
        
        \#\#\# INPUT FORMAT: 
        I will provide code snippets in the following format:...
        
        \#\#\# OUTPUT FORMAT: 
        Return your analysis ONLY in the following JSON format:
        \begin{verbatim}
{"is_malicious": <Boolean>,
"category":<Backdoor/Ransomeware/...>,
"analysis": "<Your brief reasoning>",...}
\end{verbatim}

        \#\#\#EXAMPLES: ...
        
        \#\#\#SPECIAL INSTRUCTIONS: ...
        
    \end{snapshotbox}

    
    } 
    \caption{The prompt template for LLM-based analysis.}
    \label{fig:prompt_structural}
\end{figure}

\subsection{LLM-based Semantic Analysis}
\label{sec:judge}
The final stage of the \toolname involves semantic analysis on high-fidelity contexts to determine malicious intent. This allows the system to distinguish between functionally similar but intentionally different behaviors, such as a legitimate update check versus a malicious data exfiltration. To guide the LLM in this complex task, we engineer a curated prompt template, partially illustrated in \autoref{fig:prompt_structural}, which systematically integrates four key prompting skills. 1) \textit{Role Assignment}: We instruct the LLM to assume the role of a ``Security Expert'' to activate its specialized knowledge and reasoning patterns acquired during pre-training. 2) \textit{CoT Guidance}: We prompt the LLM to follow an explicit four-step reasoning process to structure its analysis. It must first identify the sensitive operations within the slice, then trace the data flow to these operations, subsequently assess the likely intent based on this flow, and finally synthesize these findings into a conclusive judgment. This structured approach makes the LLM's reasoning process more transparent and enhances its analytical robustness. 3) \textit{In-Context Learning}: We provide several few-shot examples of benign, suspicious, and malicious code snippets along with their expected analyses. These examples provide a clear reference for the model to adhere to our required output format, and they offer concrete demonstrations of the CoT process in various scenarios. 4) \textit{Standardized JSON Output}: We require the LLM to return its findings exclusively in a uniform JSON format for machine-parsable results, including well-defined fields like \texttt{is\_malicious} and \texttt{severity}. By combining these techniques, this LLM-based semantic analysis process becomes efficient, precise, and interpretable, forming the decision-making core of the \toolname framework.
\section{Evalution}

To demonstrate the effectiveness, generalization, and scalability of \toolname, we conducted extensive experiments addressing four RQs.
\begin{itemize}[leftmargin=15pt]
    \item \textbf{RQ1: Effectiveness \& Generalization.}  
    How effective is \toolname in detecting malicious code poisoning across multiple programming languages compared to SOTA tools? Beyond Python and JavaScript, how well does \toolname generalize to other languages?

    \textbf{RQ2: Ablation Study \& Sensitivity Analysis.} 
    What are the contributions of \toolname's key components, and how sensitive is its performance to key slicing design choices?

    \item \textbf{RQ3: Impact of LLMs.}  
    How does the choice of different LLMs affect \toolname’s performance? And, to what extent does potential data contamination influence the results?

    \item \textbf{RQ4: Practicality \& Cost-effectiveness.}  
    How practical and cost-effective is \toolname for large-scale analysis, including its end-to-end cost (e.g., time and token usage), false positive rates, and its ability to identify previously undetected malicious repositories in the wild?
\end{itemize}

\subsection{Evaluation Setup}
\label{sec:setup}

\noindent\textbf{Implementation.}
We have implemented a prototype of \toolname using over 5K lines of code~(LoC) with Python and 3K LoC with Scala, excluding any third-party libraries or open-source tools.
So far, \toolname supports 5 mainstream programming languages: Python, JavaScript, Java, Go, and PHP. In our analysis workflow, we first employ Semgrep~(v1.102.0) for a rapid, initial scan of the source code to identify calls to sensitive APIs. To enable the deeper, structural analysis required for our slicing algorithm, we then leverage Joern~(v4.0.303) to uniformly convert the multi-language source code into Code Property Graphs. 
In the final semantic analysis, we selected the open-source DeepSeek-V3 model as our LLM adjudicator, which demonstrated powerful code comprehension capabilities and contributed to the cost-effectiveness of our solution. 

\noindent \textbf{Running Environment.}
All experiments were conducted on a server running Ubuntu Linux 22.04, equipped with two AMD EPYC Milan 7713 CPUs (2.0 GHz, 64 cores, 128 threads each), 512 GB RAM (8 x 64 GB modules), two NVIDIA A100 GPUs with 80 GB memory each, and four 7.68 TB NVMe SSDs (Western Digital SN640), providing a total storage capacity of 30.72 TB.

\noindent \textbf{Dataset.}
To comprehensively evaluate \toolname, we curated two datasets.

\textit{(1) Multi-Lang-Bench.} To evaluate the effectiveness of \toolname across multiple programming languages, we constructed a comprehensive dataset called \textit{Multi-Lang-Bench} by combining two sources: the public MalwareBench~\cite{li2025malwarebench} and a multi-language dataset derived from SourceFinder~\cite{md2020sourcefinder}. 
For Python and JavaScript, we adopted the widely recognized MalwareBench~\cite{li2025malwarebench}, which aggregates multiple public malicious package databases and supplements them with internal threat intelligence from Socket Security~\cite{socketdev}. We consider MalwareBench relatively reliable for our study as it has undergone peer review and includes a large-scale pre-curated dataset. Initially, this dataset contained 4,430 malicious and 11,829 neutral NPM packages, alongside 3,770 malicious and 5,495 neutral PyPI packages. To prevent evaluation bias caused by repetitive malicious templates, we performed rigorous deduplication. After removing unreviewed samples and near-duplicates, the final retained set consists of 6,444 PyPI packages (3,188 malicious and 3,256 benign) and 9,545 NPM packages (3,438 malicious and 6,107 benign).
For other languages, including Go, Java, and PHP, we constructed a dataset based on SourceFinder~\cite{md2020sourcefinder}, which initially identified 870 potentially malicious repositories from GitHub (422 for Java, 274 for PHP, and 174 for Go). To ensure dataset quality, we first excluded repositories that were no longer accessible or labeled as ``unknown'' family in SourceFinder, resulting in 746 valid repositories (265 for Java, 169 for PHP, and 119 for Go). We then conducted a thorough manual review, where three reviewers independently cross-verified metadata (e.g., README files, descriptions, commit messages) and source code to confirm the presence of malicious logic. During this process, we filtered out repositories misidentified as malicious by SourceFinder (e.g., due to heuristic errors), Android malware (unsupported by \toolname now), and low-quality repositories such as spoofing or spam that lacked actual malicious code. After this process, we retained 75 malicious samples for Java, 143 for PHP, and 80 for Go.
To construct a balanced dataset, we randomly selected an equal number of benign repositories with high stars from GitHub. These repositories were manually verified to minimize label noise. This resulted in final datasets of 150 samples for Java, 286 for PHP, and 160 for Go.
\autoref{tab:dataset_summary} summarizes the data construction process, including the original dataset size, filtered valid data, and the final number of all samples for each language.

\begin{table}[t]
\centering
\fontsize{8}{11}\selectfont
\caption{Summary of the Multi-Lang-Bench used for evaluation.}
\label{tab:dataset_summary}
\setlength{\tabcolsep}{4pt}
\begin{tabular}{|c|c|rrr|rr|rr|}
\hline
\textbf{Language} & \textbf{Source} & \textbf{Original} & \textbf{Valid} & \textbf{Reviewed} & \textbf{Malicious} & \textbf{Benign} & \textbf{Avg. LoC} & \textbf{Avg. \#Files} \\ 
\hline
Python      & MalwareBench~\cite{li2025malwarebench} & 10,143 & -    & 6,444 & 3,188 & 3.256 & 4,089 & 16.6 \\
JavaScript  & MalwareBench~\cite{li2025malwarebench} & 12,102 & -    & 9,545 & 3,438 & 6,107 & 6,118 & 28.8 \\
Java        & SourceFinder~\cite{md2020sourcefinder} & 422    & 265  & 75    & 75   & 75   & 7,560 & 79.2 \\
PHP         & SourceFinder~\cite{md2020sourcefinder} & 274    & 169  & 143   & 143  & 143   & 6,330 & 41.6 \\
Go          & SourceFinder~\cite{md2020sourcefinder} & 174    & 119  & 80    & 80   & 80    & 9,332 & 51.4 \\
\hline
\end{tabular}
\end{table}

\textit{(2) GitHub-120K\footnote{Although \toolname\ is designed to scale to millions of repositories, we focused on this subset due to the substantial storage, download, and processing costs associated with the full StarScout dataset.}.} For the large-scale study in RQ4, our dataset was derived exclusively from the open-source data provided by StarScout~\cite{he2026six}, which has identified repositories exhibiting anomalous popularity signals, making them high-probability candidates for hosting malicious or scam-related content. From the initial list of 186K unique repositories, we successfully downloaded and prepared over 120K for further analysis. This strategy allowed us to focus our large-scale evaluation on a high-risk population of repositories to discover in-the-wild threats.

\noindent \textbf{Baseline.}
To ensure a rigorous and comprehensive comparison, we systematically reviewed recent research and selected open-source baselines that are widely used in prior works~\cite{zheng2024oscar,gao2024malguard,huang2024spiderscan}.
Specifically, we select 8 representative SOTA techniques targeting the Python and JavaScript ecosystems. The rule-based detectors included Application Inspector~\cite{applicationinspector}, OSSGadget~\cite{ossgadget}, Guarddog~\cite{guarddog}, and Bandit4mal~\cite{bandit4mal} (Python only). For learning-based approaches, we selected the SAP~\cite{ladisa2023feasibility}, Amalfi~\cite{sejfia2022practical}~(JavaScript only), MalTracker~\cite{yu2024maltracker}~(JavaScrip only), and MalGuard~\cite{gao2024malguard}~(Python only). To ensure a fair comparison, we established clear and consistent criteria for determining malicious behavior. For tools like Guarddog and OSSGadget, any reported alert was sufficient to classify a package as malicious. For ApplicationInspector and Bandit4mal, we adopted a stricter and widely-used criterion, where a package was deemed malicious only if the severity and confidence were both rated as ``High''. 
For other languages~(Java, Go, PHP), we only compared \toolname with ApplicationInspector and OSSGadget due to language support limitations, and applied the same criteria as mentioned before.
\enlargethispage{1\baselineskip}

\subsection{RQ1: Effectiveness \& Generalization}
To address RQ1, we evaluated \toolname and 8 baseline tools on the Multi-Lang-Bench dataset. As summarized in \autoref{tab:effectiveness_generalization}, these results demonstrate the consistent superiority of \toolname across all 5 programming languages and ecosystems.

\noindent \textbf{Performance on Python \& JavaScript.}
As summarized in \autoref{tab:effectiveness_generalization}, \toolname demonstrated overall superior performance compared to the baseline tools across both ecosystems. On the Python dataset, \toolname achieved an F1-score of 96.30\%, marking a significant improvement of 0.80 to 74.90 percentage points over competing baselines, with MalGuard achieving the second-best F1-score of 95.50\%. Similarly, on the JavaScript dataset, \toolname led with an F1-score of 91.04\%, surpassing the baselines by margins ranging from 5.63 to 35.43 percentage points. 
While \toolname achieved the best overall performance on the JavaScript dataset, it showed lower precision (89.00\%) compared to SAP, which achieved the highest precision of 98.50\%. This discrepancy can be attributed to SAP's tendency to prioritize precision at the expense of recall, as evidenced by its significantly lower recall (46.60\%) compared to \toolname (89.00\%). In contrast, \toolname demonstrated a more balanced approach, excelling in recall and F1-score. These results highlight the robustness and high detection efficacy of \toolname, validating its ability to effectively identify malicious packages in diverse ecosystems.

\begin{table}[t]
\centering
\caption{Performance comparison of \toolname and 8 baseline tools across different programming languages. The best performance is highlighted with \colorbox{best}{\textbf{red}}, and the second-best results are highlighted with \colorbox{second}{\textbf{blue}}.}
\label{tab:effectiveness_generalization}
\resizebox{\linewidth}{!}{%
\begin{tabular}{|c|ccc|ccc|ccc|ccc|ccc|}
\hline
\textbf{\multirow{2}{*}{Tool}} & \multicolumn{3}{c|}{\textbf{Python}} & \multicolumn{3}{c|}{\textbf{JavaScript}} & \multicolumn{3}{c|}{\textbf{Go}} & \multicolumn{3}{c|}{\textbf{Java}} & \multicolumn{3}{c|}{\textbf{PHP}} \\
\cline{2-16}
 & \textbf{Pre.} & \textbf{Rec.} & \textbf{F1} & \textbf{Pre.} & \textbf{Rec.} & \textbf{F1} & \textbf{Pre.} & \textbf{Rec.} & \textbf{F1} & \textbf{Pre.} & \textbf{Rec.} & \textbf{F1} & \textbf{Pre.} & \textbf{Rec.} & \textbf{F1} \\
\hline
AppInspector~\cite{applicationinspector} & 28.2 & 17.2 & 21.4 & 48.8 & 64.7 & 55.6 & \cellcolor{second}\textbf{35.1} & \cellcolor{second}\textbf{43.0} & \cellcolor{second}\textbf{38.6} & 42.7 & \cellcolor{second}\textbf{54.7} & \cellcolor{second}\textbf{48.0} & \cellcolor{second}\textbf{50.6} & \cellcolor{second}\textbf{63.6} & \cellcolor{second}\textbf{56.4} \\
OSSGadget~\cite{ossgadget} & 48.9 & 9.0 & 15.2 & 68.9 & 66.8 & 67.8 & 34.6 & 22.8 & 27.5 & \cellcolor{second}\textbf{44.1} & 20.0 & 27.5 & 48.2 & 28.7 & 36.0 \\
Guarddog~\cite{guarddog} & 92.4 & 92.9 & 92.6 & 75.9 & \cellcolor{second}\textbf{84.3} & 79.9 & - & - & - & - & - & - & - & - & - \\
Bandit4mal~\cite{bandit4mal} & 38.8 & 18.6 & 25.1 & - & - & - & - & - & - & - & - & - & - & - & - \\
SAP~\cite{ladisa2023feasibility} & 86.2 & 53.3 & 65.8 & \cellcolor{best}\textbf{98.5} & 46.6 & 63.3 & - & - & - & - & - & - & - & - & - \\
Amalfi~\cite{sejfia2022practical} & - & - & - & 91.5 & 84.3 & \cellcolor{second}\textbf{85.4} & - & - & - & - & - & - & - & - & - \\
MalTracker~\cite{yu2024maltracker} & - & - & - & 76.6 & 66.7 & 71.3 & - & - & - & - & - & - & - & - & - \\
MalGuard~\cite{gao2024malguard} & \cellcolor{best}\textbf{97.1} & \cellcolor{second}\textbf{94.0} & \cellcolor{second}\textbf{95.5} & - & - & - & - & - & - & - & - & - & - & - & - \\
\hline
\textbf{\toolname} & \cellcolor{second}\textbf{95.7} & \cellcolor{best}\textbf{96.9} & \cellcolor{best}\textbf{96.3} & \cellcolor{second}\textbf{93.3} & \cellcolor{best}\textbf{88.9} & \cellcolor{best}\textbf{91.0} & \cellcolor{best}\textbf{95.9} & \cellcolor{best}\textbf{87.5} & \cellcolor{best}\textbf{91.5} & \cellcolor{best}\textbf{97.1} & \cellcolor{best}\textbf{88.0} & \cellcolor{best}\textbf{92.3} & \cellcolor{best}\textbf{98.5} & \cellcolor{best}\textbf{90.2} & \cellcolor{best}\textbf{94.2} \\
\hline
\end{tabular}%
}
\end{table}

\noindent \textbf{Multi-Language Generalization.}
The results in \autoref{tab:effectiveness_generalization} highlight \toolname's exceptional cross-language generalization capabilities. On the multi-language dataset, \toolname achieved an impressive F1-score of 92.51\%, significantly outperforming the baselines, ApplicationInspector and OSSGadget, which recorded F1-scores of 47.82\% and 32.37\%, respectively. Notably, this superior performance was consistent across all 5 tested languages, with \toolname maintaining F1-scores above 85\% in every case. In contrast, the baseline tools not only underperformed but also exhibited high variability in their effectiveness across different languages. These results validate that \toolname effectively abstracts away language-specific syntax, enabling its detection engine to focus on the underlying semantic behavior of the code. This capability underscores \toolname's ability to achieve language-agnostic generalization, making it a robust solution for detecting malicious code across diverse programming ecosystems.

\noindent \textbf{False Positive Analysis.} Our analysis of false positives (FPs) highlights the inherent limitations of static detection based on code semantics. In certain corner cases, developers of benign software, particularly in domains such as system administration or automated testing, may employ programming paradigms that closely resemble known malicious behaviors (e.g., dynamically constructing and executing system commands). These scenarios, where the code exhibits genuine ambiguity, present challenges for static detection. In such instances, our LLM-based analyzer may misclassify this benign yet high-risk code as malicious.

\noindent \textbf{False Negative Analysis.} The analysis of our false negatives (FNs) reveals two key areas for future improvement. First, \toolname's current design focuses exclusively on source-level analysis, rendering it unable to detect attacks leveraging malicious dependencies. For example, a package with benign source code might introduce malicious behavior by depending on a compromised or maliciously-named package—an attack vector that lies outside \toolname's current analysis scope. Second, \toolname demonstrates limitations in detecting threats involving complex, indirect execution chains. In one observed case, malicious code used a command execution call to invoke the \texttt{powershell.exe}, which then executed a \texttt{.ps1} script file containing the actual malicious payload. Here, the critical malicious logic resides outside the immediate source code being analyzed, making it challenging for the LLM to confidently classify the package as malicious. Although employing stricter prompts could potentially address such threats, this approach would likely increase false positives in legitimate use cases, highlighting the trade-off between sensitivity and precision in detection strategies. 

\subsection{RQ2: Ablation Study \& Sensitivity Analysis}
\label{sec:RQ2 Ablation Study}

\noindent \textbf{Ablation Study.} 
To understand the contributions of individual components in \toolname, we conducted an ablation study across 5 programming languages. In this study, we evaluated two variants of \toolname:

\textit{(1) w/o TPL API identification}: This variant disables the third-party API identification mechanism, which is responsible for detecting evasive threats leveraging obscure TPLs.

\textit{(2) w/o slicing}: This variant removes the slicer and processes the full file as input to the LLM, providing the model with additional benign context but without filtering irrelevant code. Unlike the file-level configuration in the pilot study (\autoref{sec:pilot_study}), which concatenates multiple files and suffers from context overflow, this variant analyzes suspicious files individually.

\begin{table}[t]
\centering
\caption{Ablation study results on the impact of different configurations across 5 programming languages. The best results for each language are highlighted in bold.}
\label{tab:ablation_all}
\resizebox{\linewidth}{!}{%
\begin{tabular}{|l|ccc|ccc|ccc|ccc|ccc|}
\hline
\textbf{\multirow{2}{*}{Configuration}} & \multicolumn{3}{c|}{\textbf{Python}} & \multicolumn{3}{c|}{\textbf{JavaScript}} & \multicolumn{3}{c|}{\textbf{Go}} & \multicolumn{3}{c|}{\textbf{Java}} & \multicolumn{3}{c|}{\textbf{PHP}} \\
\cline{2-16}
 & \textbf{Pre.} & \textbf{Rec} & \textbf{F1} & \textbf{Pre.} & \textbf{Rec} & \textbf{F1} & \textbf{Pre.} & \textbf{Rec} & \textbf{F1} & \textbf{Pre.} & \textbf{Rec} & \textbf{F1} & \textbf{Pre.} & \textbf{Rec} & \textbf{F1} \\
\hline
\textit{w/o} TPL API   & \textbf{96.6} & 94.5 & 95.6 & \textbf{93.9} & 86.5 & 90.1 & 93.2 & 85.0 & 88.9 & 97.0 & 86.7 & 91.6 & 97.7 & 88.8 & 93.0 \\
\textit{w/o} slicing & 96.6 & 91.2 & 93.8 & 93.1 & 86.9 & 89.9 & 95.8 & 85.0 & 90.1 & 95.5 & 84.0 & 89.4 & 98.4 & 88.1 & 93.0 \\
\hline
\toolname & 95.7 & \textbf{96.9} & \textbf{96.3} & 93.3 & \textbf{88.9} & \textbf{91.0} & \textbf{95.9} & \textbf{87.5} & \textbf{91.5} & \textbf{97.1} & \textbf{88.0} & \textbf{92.3} & \textbf{98.5} & \textbf{90.2} & \textbf{94.2} \\
\hline
\end{tabular}%
}
\end{table}

The results of the ablation study are shown in \autoref{tab:ablation_all}. Removing either of the core components led to a noticeable decline in detection performance.
Disabling the third-party API identification mechanism significantly reduced recall in all tested languages, highlighting its critical role in identifying evasive threats. These threats include corner-case attacks that exploit obscure TPLs to bypass scanners reliant on a predefined set of sensitive APIs. Without this mechanism, the system struggled to identify such threats, resulting in false negatives and a lower overall F1-score.
The impact of removing the slicer was even more pronounced. Using the full file as input slightly improved precision, as the additional benign context reduced misclassifications. However, this came at the cost of a substantial drop in recall, which led to a much lower F1-score overall. This demonstrates that the slicer is not merely a cost-saving measure but an essential component for reducing noise in the input. By filtering out irrelevant code, the slicer ensures that the LLM focuses on the core malicious logic, thus improving performance and achieving a superior F1-score. It is also worth noting that because this variant processes files individually, it bypasses the severe out-of-context failures observed in the pilot study's baseline. Consequently, the performance drop here is more moderate.

\noindent \textbf{Sensitivity of Slicing Strategy.} To further validate our design choices for context extraction, we conducted an extended sensitivity study on 500 samples to evaluate the call-chain depth ($k$), component combinations, and slicing granularity.
First, we evaluated the impact of the call-chain depth constraint, as shown in \autoref{tab:depth_sensitivity}. The results demonstrate a clear trade-off between semantic completeness and context bloat. Shallow expansion ($k=1$) is highly token-efficient (averaging 7,443.8 tokens) but fails to capture deeply nested cross-function behaviors, yielding a lower F1-score of 91.04\%. As the depth increases to $k=3$, the F1-score peaks at 94.18\%. However, overly deep expansion ($k=4, 5$) continuously inflates token usage (up to 11,157.6 tokens) while introducing excessive benign noise, which distracts the LLM and degrades the F1-score to 91.98\%. Thus, $k=3$ is selected as the default configuration for an optimal effectiveness-cost balance.
Second, we conducted a fine-grained ablation on the dependency components and granularity at $k=3$, as shown in \autoref{tab:granularity_components}. Comparing the full model (V6) with its sub-variants (V3, V4, V5) reveals the necessity of each dependency type. For instance, disabling Call-Chain (CC) expansion (V5) drastically reduces token usage but severely limits the context scope, dropping the F1-score to 89.41\%. Disabling Data Dependency (DD) tracing (V4) hurts recall (86.47\%), as the LLM loses visibility into the origins of sensitive variables. 
Finally, we assessed the call-chain granularity by comparing V2 (\textit{Sliced Lines}) with V6 (\textit{Whole Function}). While extracting only the strict execution lines (V2) further reduces average token consumption to 4,031.7, it severely deteriorates both precision and recall, plunging the F1-score to 86.49\%. This suggests that removing the surrounding structural contexts of a callee, including variable initializations, loop conditions, and developer comments, can undermine the semantic integrity required by LLMs for accurate reasoning.

\begin{table}[t]
\centering
\caption{Call-Depth Sensitivity Analysis ($k=1$ to $5$)}
\label{tab:depth_sensitivity}
\begin{tabular}{@{}lcccc@{}}
\hline
\textbf{Depth ($k$)} & \textbf{Precision} & \textbf{Recall} & \textbf{F1-Score} & \textbf{Avg. Tokens} \\ 
\hline
$k=1$ & 0.9348 & 0.8872 & 0.9104 & 7,443.8 \\
$k=2$ & 0.9402 & 0.9023 & 0.9209 & 9,182.9 \\
\textbf{$k=3$ (\toolname)} & \textbf{0.9521} & \textbf{0.9318} & \textbf{0.9418} & \textbf{9,613.4} \\
$k=4$ & 0.9517 & 0.9167 & 0.9339 & 10,227.4 \\
$k=5$ & 0.9504 & 0.9012 & 0.9198 & 11,157.6 \\ 
\hline
\end{tabular}
\end{table}

\begin{table}[t]
\centering
\caption{Slicing Strategy Analysis. \textmd{CC: Call-Chain, DD: Data Dependency, CD: Control Dependency.}}
\label{tab:granularity_components}
\resizebox{0.9\linewidth}{!}{%
\begin{tabular}{@{}lllcccc@{}}
\hline
\textbf{Variant} & \textbf{Components} & \textbf{Granularity} & \textbf{Precision} & \textbf{Recall} & \textbf{F1-Score} & \textbf{Avg. Tokens} \\ 
\hline
V1 & CC only & Whole Function & 0.9365 & 0.8872 & 0.9112 & 8,201.6 \\
V2 & CC+DD+CD & Sliced Lines & 0.8889 & 0.8421 & 0.8649 & 4,031.7 \\
V3 & CC+DD & Whole Function & 0.9449 & 0.9091 & 0.9266 & 9,422.7 \\
V4 & CC+CD & Whole Function & 0.9426 & 0.8647 & 0.9020 & 8,394.6 \\
V5 & DD+CD & Whole Function & 0.9268 & 0.8636 & 0.8941 & 4,747.1 \\
\hline
\textbf{V6 (Full)} & \textbf{CC+DD+CD} & \textbf{Whole Function} & \textbf{0.9521} & \textbf{0.9318} & \textbf{0.9418} & \textbf{9,613.4} \\ 
\hline
\end{tabular}}
\end{table}

\subsection{RQ3: Impact of LLMs}
\label{sec:rq3}

\noindent\textbf{Impact of Different LLMs.}  
To evaluate the impact of different LLMs on the performance of \toolname, we conducted experiments using several well-known general-purpose LLMs, including GPT-4o, GPT-3.5 Turbo, Claude 4 Sonnet, Gemini 3 Pro, and DeepSeek-V3. 
The results, summarized in \autoref{tab:llm_comparison}, show that the performance of different LLMs varies across programming languages, with no single model consistently dominating in all cases. For instance, GPT-4o demonstrates some advantages in Python, JavaScript, and Go, achieving the highest F1 scores in these languages. However, the improvements over other models, such as DeepSeek-V3 or Claude 4 Sonnet, are not always substantial. On the other hand, DeepSeek V3 achieves competitive or even superior results in PHP and Go, while maintaining a balanced performance across all languages.  
Considering both performance and inference costs, DeepSeek V3 emerges as a practical choice for large-scale analysis. While high-performing models like GPT-4o offer marginally better results in certain cases, their significantly higher computational costs may limit their feasibility in large-scale analysis. 

\begin{table}[t]
\centering
\caption{Comparison of \toolname's performance using different LLMs across languages. The best results for each metric are highlighted with \colorbox{best}{\textbf{red}}, and the second-best results are highlighted with \colorbox{second}{\textbf{blue}}.}
\label{tab:llm_comparison}
\resizebox{\linewidth}{!}{%
\begin{tabular}{|l|ccc|ccc|ccc|ccc|ccc|}
\hline
\textbf{\multirow{2}{*}{LLM}} & \multicolumn{3}{c|}{\textbf{Python}} & \multicolumn{3}{c|}{\textbf{JavaScript}} & \multicolumn{3}{c|}{\textbf{Go}} & \multicolumn{3}{c|}{\textbf{Java}} & \multicolumn{3}{c|}{\textbf{PHP}} \\
\cline{2-16}
 & \textbf{Pre.} & \textbf{Rec.} & \textbf{F1} & \textbf{Pre.} & \textbf{Rec.} & \textbf{F1} & \textbf{Pre.} & \textbf{Rec.} & \textbf{F1} & \textbf{Pre.} & \textbf{Rec.} & \textbf{F1} & \textbf{Pre.} & \textbf{Rec.} & \textbf{F1} \\
\hline
GPT-3.5 Turbo & \cellcolor{second}\textbf{95.7} & 95.7 & 95.7 & \cellcolor{best}\textbf{97.1} & 84.2 & 90.2 & 95.2 & \cellcolor{second}\textbf{90.9} & \cellcolor{second}\textbf{93.0} & 88.2 & \cellcolor{second}\textbf{93.8} & 90.9 & \cellcolor{second}\textbf{97.9} & 82.1 & 89.3 \\
GPT-4o & \cellcolor{best}\textbf{98.6} & \cellcolor{best}\textbf{97.0} & \cellcolor{best}\textbf{97.8} & \cellcolor{second}\textbf{94.3} & \cellcolor{best}\textbf{91.9} & \cellcolor{best}\textbf{93.1} & \cellcolor{second}\textbf{95.5} & \cellcolor{best}\textbf{91.3} & \cellcolor{best}\textbf{93.3} & 95.2 & 93.0 & \cellcolor{best}\textbf{94.1} & 96.0 & \cellcolor{second}\textbf{87.3} & \cellcolor{second}\textbf{91.4} \\
Claude 4 Sonnet & 95.2 & 95.2 & 95.2 & 94.1 & 86.8 & 90.3 & 92.9 & 89.7 & 91.2 & \cellcolor{best}\textbf{97.6} & 90.9 & \cellcolor{best}\textbf{94.1} & 94.0 & 83.9 & 88.7 \\
Gemini 3 Pro & 86.9 & 95.2 & 90.9 & 78.7 & \cellcolor{second}\textbf{91.3} & 84.5 & 82.0 & 89.3 & 85.5 & 83.9 & \cellcolor{best}\textbf{94.0} & 88.7 & 85.1 & 87.0 & 86.0 \\
DeepSeek-V3 & \cellcolor{second}\textbf{95.7} & \cellcolor{second}\textbf{96.9} & \cellcolor{second}\textbf{96.3} & 93.3 & 88.9 & \cellcolor{second}\textbf{91.0} & \cellcolor{best}\textbf{95.9} & 87.5 & 91.5 & \cellcolor{second}\textbf{97.1} & 88.0 & \cellcolor{second}\textbf{92.3} & \cellcolor{best}\textbf{98.5} & \cellcolor{best}\textbf{90.2} & \cellcolor{best}\textbf{94.2} \\
\hline
\end{tabular}%
}
\end{table}

\noindent\textbf{Data Contamination Analysis.}
A potential concern in using LLMs for malicious package detection is the possibility of data contamination, where some malicious packages in the evaluation set may already exist in the LLM's training data. To address this issue, we leveraged the Multi-Lang-Bench dataset, whose packages were all released before May 2024, as the subset of data published \textit{before} the LLM training cut-off date. This ensures that these packages were released before the publication dates of both DeepSeek-V3 and GPT-4o. Additionally, we randomly collected 260 malicious packages released after January 2025 from the dataset maintained by DataDog~\cite{datadogsamples}~(130 for Python and JavaScript each). This specific date ensures that these packages were published after the release of both DeepSeek-V3 and GPT-4o, guaranteeing that none of these malicious packages appear in the training data of either of the two LLMs.
The performance comparison, shown in \autoref{tab:contamination_analysis}, reveals that while there are slight performance differences between the two subsets, the results remain robust overall. For packages released \textit{before} the cut-off date, there is a marginal improvement in precision and recall, which could suggest potential effects of data contamination, as these packages might have been seen during the LLM's training phase. However, for packages released \textit{after} the cut-off date, the system still demonstrates strong performance, indicating that \toolname effectively generalizes to unseen malicious packages. These findings suggest that while data contamination may introduce a slight advantage, its impact on the evaluation results is relatively acceptable and does not undermine the broader conclusions about \toolname's generalization capabilities.

\noindent\textbf{Reliability of LLM-based TPL Identification.}
A potential concern in LLM-assisted analysis is hallucination, particularly during the identification of sensitive third-party APIs. To systematically evaluate the output quality of our LLM-based TPL extractor, we conducted a manual verification study. Specifically, we randomly sampled and manually reviewed 50\% (347 out of 694) of the TPL APIs flagged as sensitive by the LLM during our multi-language evaluation. Our manual inspection revealed only 4 false positives within this sample. This low error rate suggests that, for this well-defined semantic extraction task, the LLM can produce accurate and stable outputs when guided by explicit function signatures, immediate implementation bodies, and structured prompt engineering.

\begin{table}[t]
\centering
\caption{Performance comparison on packages released before and after the LLM training cut-off date for different LLMs. The increase is highlighted with \colorbox{best}{\textbf{red}}, and the decrease is highlighted with \colorbox{second}{\textbf{blue}}.}
\label{tab:contamination_analysis}
\resizebox{0.75\linewidth}{!}{%
\begin{tabular}{|c|c|ccc|ccc|}
\hline
\textbf{\multirow{2}{*}{LLM}} & \textbf{\multirow{2}{*}{Subset}} & \multicolumn{3}{c|}{\textbf{Python}} & \multicolumn{3}{c|}{\textbf{JavaScript}} \\
\cline{3-8}
 & & \textbf{Pre.} & \textbf{Rec.} & \textbf{F1} & \textbf{Pre.} & \textbf{Rec.} & \textbf{F1} \\
\hline
\multirow{3}{*}{DeepSeek V3} 
 & Before Cut-off & 95.7 & 96.9 & 96.3 & 93.3 & 88.9 & 91.0 \\ 
 & After Cut-off  & 95.7 & 92.6 & 94.1 & 89.2 & 92.5 & 90.8 \\ 
 & \textbf{Change}         & \textbf{\cellcolor{second}{0.0\%}} & \textbf{\cellcolor{best}$\downarrow$~4.3\%} & \textbf{\cellcolor{best}$\downarrow$~2.2\%} & \textbf{\cellcolor{best}$\downarrow$~4.1\%} & \textbf{\cellcolor{second}$\uparrow$~3.6\%} & \textbf{\cellcolor{best}$\downarrow$~0.2\%} \\ 
\hline
\multirow{3}{*}{GPT-4o} 
 & Before Cut-off & 98.6 & 97.0 & 97.8 & 94.3 & 91.9 & 93.1 \\ 
 & After Cut-off  & 95.3 & 91.0 & 93.1 & 88.0 & 90.4 & 89.2 \\ 
 & \textbf{Change}         & \textbf{\cellcolor{best}$\downarrow$~3.3\%} & \textbf{\cellcolor{best}$\downarrow$~6.0\%} & \textbf{\cellcolor{best}$\downarrow$~4.7\%} & \textbf{\cellcolor{best}$\downarrow$~6.3\%} & \textbf{\cellcolor{best}$\downarrow$~1.5\%} & \textbf{\cellcolor{best}$\downarrow$~3.9\%} \\ 
\hline
\end{tabular}%
}
\end{table}

\subsection{RQ4: Practicality and Cost-effectiveness}  
To evaluate the cost-effectiveness of \toolname, we randomly selected 2,168 repositories from the GitHub-120K dataset for analysis. These repositories contain a total of 118,200 source files, amounting to over 300 million tokens. A naive file-by-file LLM analysis of such a dataset would be prohibitively expensive. However, as shown in \autoref{tab:scalability_reduction}, \toolname achieved a substantial reduction in workload. The initial sensitive API identification effectively filtered out 83.5\% of the files~(from 118,200 to 19,477), significantly narrowing the analysis scope. The remaining suspicious files were then refined using the hybrid slicing technique, which further reduced the content to be analyzed. Ultimately, the token cost of the LLM-based semantic analysis was reduced to just 19.2 million, which means a 94.0\% reduction compared to file-by-file analysis. 
In terms of time cost, incorporating parsing and slicing techniques did not significantly increase the overall analysis time. As shown in \autoref{tab:scalability_reduction}, the parsing step took only 40.6 seconds per project, while the slicing stage added another 53.3 seconds. These steps accounted for less than 100 seconds in total, an acceptable overhead given the cost reduction in the subsequent LLM-based analysis stage. These results demonstrate that \toolname effectively narrows the scope of analysis without imposing substantial overhead, making \toolname a practical solution for large-scale repo-level analysis.

\begin{table}[t]
\centering
\caption{Token cost reduction, scalability, and end-to-end time analysis across 2168 projects.}
\label{tab:scalability_reduction}
\resizebox{\textwidth}{!}{%
\begin{tabular}{|c|c|r|r|r|r|r|r|r|}
\hline
\multirow{2}{*}{\textbf{Context}} & \multirow{2}{*}{\textbf{\#Projects}} & \multirow{2}{*}{\textbf{\#Files}} & \multirow{2}{*}{\textbf{Total Tokens}} & \multirow{2}{*}{\textbf{Cost (\$)}} & \multirow{2}{*}{\textbf{Savings}} & \multicolumn{3}{c|}{\textbf{End-to-End Time (s)}} \\ 
\cline{7-9}
 &  &  &  &  &  & \textbf{Parsing} & \textbf{Slicing} & \textbf{LLM} \\ 
\hline
All Source Files & \multirow{3}{*}{2168} & 118,200 & 319,459,902 & 86.25 & {-} & - & - & 252.4 \\ 
\cline{1-1} \cline{3-9}
Suspicious Files &  & 19,477 & 118,034,161 & 31.87 & $\downarrow$~63.0\% & 40.6 & - & 83.7 \\ 
\cline{1-1} \cline{3-9}
Suspicious Slices &  & {N/A} & 19,221,564 & 5.19 & $\downarrow$~94.0\% & 40.6 & 53.3 & 46.2 \\ 
\hline
\end{tabular}%
}
\end{table}

To evaluate the real-world scalability and effectiveness of \toolname, we deployed it on the GitHub-120K dataset, comprising 7.3 million files. The entire analysis pipeline consumed 1,255 million tokens at a total cost of approximately \$338, demonstrating the economic feasibility of our approach at scale. \toolname initially reported 1,553 repositories as potentially malicious, which were subsequently subjected to a rigorous manual review. This review was independently conducted by two security researchers with over 5 years of experience, with any disagreements resolved by a senior expert possessing 7 years of experience. To ensure precision, repositories were only confirmed as malicious if they exhibited clear, undeniable malicious intent. This review process filtered out 502 false positives. Although this yields a False Discovery Rate (FDR) of 32.2\%, the overall False Positive Rate (FPR) across the 120K analyzed repositories remains remarkably low. Further investigation into these false positives revealed that they primarily stem from ``gray-ware'' or benign utilities whose behaviors closely mimic malicious actions (e.g., legitimate remote administration tools or system monitors executing shell commands). 
We also excluded 487 repositories that explicitly self-identified in their README files as being for penetration testing, security research, or malware collection. This process yielded a final set of 564 confirmed, previously unknown malicious repositories. Specifically, we identified 200 repositories in C/C++, 174 in Python, 56 in JavaScript, 31 in Java, and 21 in Go, with the remainder in other environments. The dominant malicious patterns included credential exfiltration (e.g., token grabbers and injectors, 73 cases) and direct payload execution (71 cases), alongside covert cryptocurrency miners and deceptive malicious forks. Following responsible disclosure guidelines, all 564 confirmed repositories were reported to the GitHub security team. This large-scale study demonstrates that \toolname is not only capable of scaling cost-effectively to analyze massive codebases but is also highly effective at uncovering multi-language threats that evade traditional security measures.
\section{Discussion}
\label{sec:discussion}

\textbf{Limitations.}
\toolname has several limitations arising from practical and methodological constraints. First, the current implementation of \toolname is affected by the maturity of the underlying static analysis toolchain. Although its LLM-based semantic analysis is language-agnostic in principle, the slicing stage requires language-aware static-analysis support. Therefore, extending \toolname to a new language mainly requires integrating a suitable parsing or slicing backend and curating an initial set of built-in sensitive APIs. While our prototype uses a Joern-based CPG backend, other mature multi-language analysis frameworks, such as CodeQL~\cite{codeql} and YASA~\cite{wang26yasa}, could also be incorporated to provide parsing, dependency analysis, or slicing capabilities.
Second, \toolname is designed for multi-language but not cross-language attacks. It cannot trace data or control flow in language boundaries—e.g., when a Python script invokes a C library, which creates a significant blind spot for advanced attacks where malicious payloads are concealed in native extensions. 
Third, to ensure scalability across millions of repositories, our pre-filtering stage deliberately omits complex intra- and inter-procedural alias resolution. While our high overall recall (averaging over 90\%) empirically demonstrates that this trade-off is acceptable for current real-world threats, it inherently creates a blind spot for sophisticated attacks that exclusively hide sensitive operations behind deep, multi-hop alias chains. Finally, our reliance on LLMs introduces a new attack surface.
Finally, our reliance on LLMs introduces a new attack surface. The LLM-based analyzer is susceptible to prompt injection attacks, where embedded instructions in code could manipulate the model’s reasoning. For example, attackers could embed deceptive comments or misleading strings to trick the model into classifying malicious code as benign. 

\noindent\textbf{Future Work.}
The modular architecture of \toolname opens up opportunities for future enhancements and applications. One key direction is to expand its applicability beyond the current programming languages and repositories to encompass other ecosystems, such as VSCode extensions, MCP servers, agent skills, and other domain-specific software ecosystems. Additionally, the extensibility of \toolname enables support for more programming languages and addressing more sophisticated threats, such as C/C++, Ruby, etc. 
Another important area of improvement is incremental analysis. For large-scale codebases, analyzing every file during each scan can be computationally expensive and time-consuming. Instead, \toolname can focus on analyzing changes introduced in pull requests (PRs) or commits, significantly reducing the overhead while maintaining security coverage. 
Finally, to maximize the real-world impact of \toolname, we are actively collaborating with the VirusTotal team to integrate \toolname into their platform.
\section{Related Work}
\label{sec:related_work}
Software supply chain security has become a critical research focus due to the increasing prevalence of attacks targeting package registries and open-source repositories.

\noindent \textbf{Attack Vectors in Software Supply Chain}
Recent research has extensively examined the diverse and evolving attack vectors in software supply chains. Gu et al.~\cite{gu2023package} systematically investigated six major software registry ecosystems, identifying twelve potential attack vectors, including six novel ones. Their work highlights how adversaries exploit inconsistencies in registries, mirrors, and clients to distribute malicious code. Similarly, Zahan et al.~\cite{zahan2022weaklinks} analyzed metadata from over 1.63 million npm packages, proposing six signals of security weaknesses, such as install scripts and expired maintainer email domains.  
To structure this complex threat landscape, Ladisa et al.~\cite{ladisa2023sokattack} proposed a comprehensive taxonomy of 107 unique attack vectors, covering all stages of the supply chain and linking them to real-world incidents and safeguards. Neupane et al.~\cite{neupane2023typosquatting} further expanded on specific threats, categorizing 13 distinct mechanisms of package confusion attacks that adversaries use to mislead developers, moving beyond traditional typosquatting.  
Finally, Wermke et al.~\cite{wermke2023opensource} explored the challenges companies face when integrating open-source components, such as the difficulty of auditing dependencies and the lack of dedicated resources for managing supply chain risks. Their findings emphasize the importance of improving security practices and fostering a healthier open-source ecosystem.

\noindent \textbf{Mitigating Malicious Package in PyPI/NPM Ecosystems.}
In response to these threats, a diverse array of detection techniques has been developed. Established approaches leverage \textit{rule-based program analysis}\cite{duan2021maloss,li2023malwukong}, using either static analysis engines like CodeQL for semantic signature matching\cite{gobbi2023codeql} or dynamic analysis in sandboxes to monitor runtime behavior~\cite{zheng2024oscar, cheng2024donapi, tanzir2025dysec}. However, a critical barrier for these tools is their often unacceptably high false positive rate in real-world deployments~\cite{vu2023badsnakes}. To improve generalization, \textit{learning-based methods} learn patterns from features extracted from package metadata~\cite{sajal2024metadata,haya2025mlpypi} or source code~\cite{sejfia2022practical, simone2022feasibility, ladisa2023feasibility}. While more flexible, their performance is fundamentally tied to hand-crafted features that may fail to capture the semantic nuances of novel attacks. The advent of \textit{LLMs} has opened a new frontier. However, their application has often been limited to auxiliary tasks like sensitive API identification~\cite{gao2024malguard, huang2024spiderscan}, feature generation~\cite{wang2025malpacdetector}, or cross-language data augmentation~\cite{yu2024maltracker}. Studies that do use LLMs as direct judges have highlighted their potential but also underscored the prohibitive costs and context window limitations~\cite{zahan2025socketai,di2024posterllm}.
While these works provide valuable insights into malicious package detection, \toolname addresses multi-language generalization challenges and context limitations.

\noindent \textbf{Mitigating Malicious Behavior in GitHub Repositories.}
Beyond package registries, GitHub has been identified as a significant reservoir for malicious source code~\cite{md2024educational, md2020sourcefinder, kKaminskiDS24study}. However, existing detection methods for this domain predominantly operate at a coarse-grained, non-code level. These approaches typically rely on identifying malicious projects through repository similarity and clustering based on metadata and file structure~\cite{rokon21Repo2Vec}, or by training models on project-level metadata such as README content or anomalous social signals like fake stars~\cite{he2026six}. While computationally efficient for flagging certain types of malicious activity, these methods fundamentally bypass deep, semantic inspection of the source code itself. This leaves a critical blind spot for novel or stealthy malware that does not exhibit obvious structural or social anomalies.
In contrast, \toolname addresses their limitations by enabling fine-grained semantic analysis of source code, leveraging LLMs to detect malicious behaviors in language-heterogeneous Github codebases.
\section{Conclusion}
\label{sec:conclusion}
In this paper, we propose \textsc{MalTotal}, a novel framework for scalable and language-agnostic malicious code detection in large-scale codebases. By integrating LLM-assisted semantic reasoning with a hybrid semantic slicing strategy, \textsc{MalTotal} effectively identifies and analyzes malicious behaviors across diverse programming languages while reducing token consumption by 94.0\%. Extensive evaluations demonstrate its superiority over 8 state-of-the-art baselines, achieving an average F1-score of 93.1\% across 5 programming languages. Furthermore, \textsc{MalTotal} successfully scaled to analyze 120K GitHub repositories, identifying 564 previously unknown malicious repositories with a total cost of just \$338. These results validate \textsc{MalTotal}'s effectiveness, cost-efficiency, and scalability, highlighting its practical potential to mitigate large-scale code poisoning attacks.

\section*{Acknowledgment}
This work was supported in part by the National Natural Science Foundation of China (grants No. 62502168, 62572209), and by the Hubei Provincial Key Research and Development Program (grant No. 2025BAB057).

\section*{Statement on the Use of AI Tools}
We confirm that this work involved the use of generative AI tools for polishing the language of the paper as well as for LLM-assisted coding in non-critical tasks. No AI-generated content contributed to the scientific or technical originality of the paper. All use of AI tools complies with ACM's policies on authorship and the use of generative AI technologies.

\section*{Data Availability Statement}
We have made the implementations of \toolname and experimental data publicly accessible at 
\url{https://github.com/security-pride/MalTotal}.

\bibliographystyle{ACM-Reference-Format}
\bibliography{reference}

\end{document}